\documentclass{rist}
\usepackage{indentfirst}

\begin{document}

\title{Experiments with a PCCoder extension}

\author{Mircea-Dan Hernest\\\url{danhernest@gmail.com}}

\date{Romanian Institute of Science and Technology\\3 Virgil Fulicea, 400022 Cluj-Napoca, Romania}

\maketitle

\begin{abstract}
  Recent research in synthesis of programs written in some Domain Specific Language (DSL) by means of neural networks
  from a limited set of inputs-output correspondences such as DeepCoder \cite{DeepCoder} 
  and its PCCoder \cite{PCCoder} 
  reimplementation/optimization proved the efficiency of this kind of approach
  to automatic program generation in a DSL language that although limited in scope is universal in the sense that
  programs can be translated to basically any programming language.

  We experiment with the extension of the domain specific language (DSL) of DeepCoder/PCCoder with symbols {\em IFI} and {\em IFL}
  which denote functional expressions of the {\em If} ramification (test) instruction for types Int and List.
  We notice an increase (doubling) of the size of the training set, the number of parameters of the trained
  neural network and of the time spent looking for the program synthesized from limited sets of inputs-output correspondences.

  The result is positive in the sense of preserving the accuracy of applying synthesis on randomly generated
  test sets.
\end{abstract}

\section{Introduction}

The task of automatically finding a program in some underlying programming language that satisfies the user intent
expressed in the form of some specification, i.e., program synthesis, has been considered ‘the holy grail’ of
Computer Science ever since the inception of Artificial Intelligence (AI) in the 1950’s. It was Alonzo Church who
defined the problem to synthesize a circuit from mathematical requirements during the Summer Institute of Symbolic
Logic at Cornell University in 1957. With the revival of artificial Neural Networks during the last decade the issue
of applying deep learning techniques to such an old problem finally percolated in the last two years, as illustrated
by the papers listed in the bibliography section \cite{NBGPS, PAGR, EGNPS, WEA, NGDS, NPS, SDN}.
This basically started with the initial success of Balog et. al. \cite{DeepCoder} from Google Brain.
The reimplementation and adaptation of Zohar and Wolf \cite{PCCoder} added further depth by using a second neural network
for pruning and thus extending the length of generated programs.

In their 2017 survey \cite{PS}, Gulwani, Polozov and Singh present a review of state of the art techniques for program synthesis
that include in Section 6.4 the approach called Neural Program Synthesis which employs Neural Networks in one of two ways:
either new neural architectures learn the behavior of a program that is consistent with a given set of input-output examples
or the neural systems do perform effective program synthesis by returning an interpretable program that matches the desired
specification behavior. The first approach is also called ‘program induction’ and comes with the shortcoming that no
interpretable model of the learnt program is generated while large computational resources and several thousands of input-output
examples per synthesis task are required.

We are thus concerned with the second approach of which Flash Fill is a successful
example that has already been implemented in Microsoft Excel. Given the Flash Fill domain specific language (DSL),
the Flash Fill neural system learns a generative model of programs in the DSL which is conditioned on the input-output examples.
The neural system is made from two networks: the I/O encoder and the R3NN  (Recursive-Reverse-Recursive Neural Network)
which incrementally synthesizes a program in the DSL given a continuous representation of the input-output examples.

Other, non-neural approaches to program synthesis include: enumerative search, constraint solving,
stochastic search, and deduction-based programming by examples (see, e.g., \cite{RGPS}).

\section{Deep Coder - PC Coder}

In their paradigm-shifting approach \cite{DeepCoder}, Balog et al. achieve the goal of integrating neural network
architectures with search-based techniques (rather than replace them) for the objective of program induction:
they illustrate the use of a corpus of program induction problems in learning strategies that generalize across problems.
They manage to define a programming language that is expressive enough to include real-world programming problems
while being high-level enough to be predictable from inputs-output examples.

They generate models for mapping sets
of inputs-output examples to program properties and they perform experiments that show an order of magnitude speedup
over standard program synthesis techniques, which makes their approach feasible for solving problems of similar
difficulty as the simplest problems that appear on programming competition websites.

In their PCCoder reimplementation \cite{PCCoder}, Zohar and Wolf optimize and extend the performance of DeepCoder
by employing a stepwise approach to the program synthesis problem. Given the current state, the main neural network
directly predicts the next statement, including both the function (operator) and parameters (operands). A beam search
is then performed, based on the network's predictions, reapplying the neural network at each step. The more accurate
the neural network, the less programs one needs to include in the search before identifying the correct solution.
Since the number of variables increases with the program's length, some of the variables in memory need to be discarded.
Therefore a second network is trained to predict the variables that are to be discarded. Training the new network does
not only enable to solve more complex problems, but serves also as an auxiliary task that improves the generalization
of the statement prediction (main) network.

The DSL from \cite{DeepCoder,PCCoder} is a small purely functional language with basic types {\em Bool}, {\em Int} and
{\em List} where the latter are vectors of at most 20 {\em Int} and the functions are divided into three catgories:
first-order, higher-order and lambdas. Each line of instruction is given by a function name followed by arguments
0,1,2, .. where each natural represents the n-th (previous) instruction (result) or one of the inputs:
0 identifies the first input, 1 the second input (if it exists), 2 the third input or the result of the first instruction
(if our program only takes two inputs). Thus no constants are allowed in their language, as these natural literals
are reserved for positions in the instruction stack. The outcome of the program is
the result of the application of the last instruction in the stack.

The DSL of PCCoder is yet the same as the one of DeepCoder and currently cannot express solutions to many problems.
Complex problems require more complex algorithmic solutions like dynamic programming and search, which are currently
beyond reach for such DSL.

It is therefore important to attempt the extension of the DSL with branching and loop constructs and investigate
how this would complicate the task of program synthesis in terms of the increase in search complexity. We
previously demonstrated that at least allowing more function symbols appears to be feasible, without much
complexity increase, see Section \ref{sc_MNIST}.

We experimented with the addition of the branching command {\em If} expressed functionally for types Int and List.
We thus extended the DSL with symbols {\em IFI} and {\em IFL} which denote functional expressions of the {\em If}
ramification (test) instruction for types Int and respectively List:\\[12pt]\hspace*{-1cm}
\begin{minipage}[l]{20cm}
\begin{verbatim}
IFI = Function('IFI', lambda f, n, x, y: 
    x if f(n) else y, (FunctionType(INT, BOOL), INT, INT, INT), INT)

IFL = Function('IFL', lambda f, n, x, y: 
    x if f(n) else y, (FunctionType(INT, BOOL), INT, LIST, LIST), LIST)
\end{verbatim}
\end{minipage}\\[12pt]
We noticed an increase in the size of the training set $T_n$ (basically a doubling in size) for programs of
length at most $n$ (e.g., $n=5$) and correspondingly an increase (doubling) of the number of parameters
of the neural network $R_n$ trained from $T_n$. All this also brings a doubling of the execution time of
the search script for solutions from examples. The maximum length of a program was not affected, as networks
could be generated even for $n\ge 20$.

We also noticed the generalization capacity of the result in the sense that, e.g., the network $R_5$ has a success
rate of $\ge 90\%$ for a randomly generated test set of programs of length $6$, a success rate of
$\ge 70\%$ for programs of length $8$ and $\approx 50\%$ for programs of length $14$. In general, the
network $R_n$ has a $99\%$ success rate for programs of length at most $n$, also for the extended DSL. 

\section{MNIST classification}
\label{sc_MNIST}

We first demonstrated a simpler (from a complexity viewpoint) extension of the DSL, namely with function symbols
for Python implemented functions. This may have at most a linear impact on the complexity, as the new symbol
may be included in the programs of a certain maximal length generated for training the network later used
in the program synthesis search for a program compatible with a certain given set of inputs-output pairs.

We exemplify with a function that is complex enough in order to be interesting but also has a fairly simple
implementation in Python (using the {\em sklearn} library), namely a function for MNIST classification.

We thus extended the DSL of DeepCoder/PCCoder with a function symbol {\em MNIST} which takes as input
a list of $8$ integers between $0$ and $255$ and returns an integer between $0$ and $9$ or $10$ as error.
Each of the $8$ integers represents a line in the 8x8 matrix of the input drawing, with the integer
value between $0$ and $255$ coding the columns marked with $1$ in the binary representation where
a point is drawn (on that line).

We can thus take advantage of the Python library {\em sklearn}'s
{\em datasets.load\_digits()} training dataset for digits classification where each datapoint is a
8x8 image of a digit (we also transform the standard values $0$ to $16$ to a binary $0$/$1$).

Thus for MNIST we use 8x8 images with binary pixel values (1 for black and 0 for white)
which come at input as one-byte unsigned integer arrays of length 8: the one-byte translates
to integers in the range [0,.., 255].

Here is the code that we add to {\em dsl/impl.py}:

\begin{verbatim}
from sklearn.datasets import load_digits
from sklearn.ensemble import RandomForestClassifier
import numpy as np

def bin(n):
    if n<8:
        return 0
    else:
        return 1

def bin2dec(a):
    r=0
    for i in range(8):
       r+=a[i]*(2**i)
    return r

digits = load_digits()
data=digits.data
labels=digits.target
m=data.shape[0]
n=data.shape[1]

for i in range(m):
  for j in range(n):
     data[i][j]=bin(data[i][j])

data=np.reshape(data,(m,8,8))
d=[[0 for x in range(8)] for y in range(m)]

for i in range(m):
    for j in range(8):
        d[i][j]=bin2dec(data[i][j])

clf_rf = RandomForestClassifier()
clf_rf.fit(d, labels)

def mnist(tx):
    if len(tx)!=8:
       return 10
    for k in range(8):
       if tx[k]<0 or tx[k]>255:
          return 10
    print("predict BEGIN")
    y_pred_rf = clf_rf.predict([tx])
    print("predict END")
    return int(y_pred_rf[0])
\end{verbatim}

In the file \emph{dsl/impl.py} we also add the following line in the section outlined by ``\# first-order functions'' :
\begin{verbatim}
MNIST = Function('MNIST', mnist, LIST, INT)
\end{verbatim}

In the file \emph{dsl/constraint.py} we also need to replace
\begin{verbatim}
elif f in [impl.MINIMUM, impl.MAXIMUM]:
        # list constrained to int constraint
        return [ListConstraint(int_constraints=[constraint] * L)]
\end{verbatim}
with (just add `impl.MNIST' to the list) 
\begin{verbatim}
elif f in [impl.MINIMUM, impl.MAXIMUM, impl.MNIST]:
        # list constrained to int constraint
        return [ListConstraint(int_constraints=[constraint] * L)]
\end{verbatim}

We only need to generate DSL programs of length at most 2, as our target is a program of
length 1 (which simply applies \texttt{mnist} on its LIST input to return an INT).
After training the two networks on the set of generated programs, we run
\emph{scripts/gen{\_}programs.py} on the following test set of ten correspondences of
label (one of 0,..,9) to image (array of 8 integers from 0 to 255):

\begin{verbatim}
{"examples": [{"inputs": [24, 60, 100, 100, 100, 36, 52, 24], 
"output": 0}, {"inputs": [24, 56, 24, 28, 24, 24, 24, 56], 
"output": 1}, {"inputs": [48, 56, 60, 48, 28, 14, 60, 112], 
"output": 2}, {"inputs": [24, 22, 24, 24, 48, 96, 100, 56], 
"output": 3}, {"inputs": [16, 16, 8, 104, 44, 60, 48, 16], 
"output": 4}, {"inputs": [12, 60, 60, 28, 32, 96, 48, 60], 
"output": 5}, {"inputs": [24, 24, 12, 12, 12, 60, 120, 56], 
"output": 6}, {"inputs": [120, 96, 48, 60, 28, 8, 12, 4], 
"output": 7}, {"inputs": [28, 60, 44, 56, 24, 60, 100, 60], 
"output": 8}, {"inputs": [12, 60, 60, 52, 60, 96, 48, 28], 
"output": 9}]}
\end{verbatim}

It is sufficient to only search for programs of length 1 and we quickly (a few seconds)
obtain the following DSL program:

\begin{verbatim}
{"result": "LIST|MNIST,0", "num_steps": 27, "time": 2.6361, 
"beam_size": 200, "num_invalid": 3, "width": 20}
\end{verbatim}

\section{Extension with ramification instructions}

Since the DSL of DeepCoder/PCCoder is essentially functional, in order to add the ramification (test) instruction
{\em If-Then-Else} we had to consider its functional versions {\em IFI} and {\em IFL} for types {\em Int} and
{\em List} respectively.

The test value is of type {\em Bool} and is obtained as function of an {\em Int} which
in the DSL can be one of the three comparisons to $0$ or an even/odd test:
\begin{verbatim}
EQ0 = Function('=0', eqZero, INT, BOOL)
GT0 = Function('>0', gtZero, INT, BOOL)
LT0 = Function('<0', ltZero, INT, BOOL)
EVEN = Function('EVEN', isEven, INT, BOOL)
ODD = Function('ODD', isOdd, INT, BOOL)
\end{verbatim}

Notice that {\tt EQ0} was not part of the original DSL, we added it for increasing the expressivity of the language.
The Python functions are defined as expected (but not using lambdas as in \cite{PCCoder}, due to technical problems)

\begin{verbatim}
def eqZero(x):  return bool(x==0)
def gtZero(x):  return bool(x>0)
def ltZero(x):  return bool(x<0)
def isEven(x):  return bool(x%2==0)
def isOdd(x):  return bool(x%2==1)
\end{verbatim}

We present the steps of operation of the PCCoder implementation \cite{GPCC}.
First a sufficiently large training set is generated (and eventually also a smaller test set)
with the option of loading a cache of previously generated training set (of programs of smaller maximal length).
\begin{verbatim}
python -m scripts.gen_programs --num_train=100000 --num_test=500 
--train_output_path=train_dataset --test_output_path=test_dataset 
--max_train_len=12 --test_lengths="5 9" --num_workers=20
\end{verbatim}

Here \texttt{num\_train} is the raw number of programs generated before eliminating equivalent
programs, \texttt{num\_test} is the number of test programs for each length given to
\texttt{test\_lengths}; \texttt{num\_workers} is the number of processes tasked to
concurrently execute the workload, \texttt{train\_dataset} is the file name of the train
dataset and \texttt{test\_dataset} is the file name of the test dataset. Also \texttt{max\_train\_len}
is the maximal length of a generated program and there is also the optional parameter \texttt{cache}
that gives the file name of a previously generated train dataset. The script will first find the
maximal length of a program from the cache database and proceed with further generation only
if this is strictly lower than \texttt{max\_train\_len}.

The parameter \texttt{num\_train} should not be larger than the number of all possible raw generated
programs and therefore has to be fixed for small \texttt{max\_train\_len} values such as $1$ and $2$.
It is possible to find the maximal values for $1$ and $2$ by running the script with a large
\texttt{num\_train} and noting the number of raw programs generated before the script stops,
e.g., the following line is written in the preamble of {\em scripts/gen\_programs.py} for the
DSL without {\em IFI} and {\em IFL}
\begin{verbatim}
KNOWN_TRAIN_SIZES = {1: 44, 2: 2561}
\end{verbatim}
whereas for the richer DSL with {\em IFI} and {\em IFL} we have more possibilities:
\begin{verbatim}
KNOWN_TRAIN_SIZES = {1: 123, 2: 15000}
\end{verbatim}

Already for \texttt{max\_train\_len}=3 many more programs of length 3 can be generated in the richer
DSL; note that it is recommended to include virtually all programs of length 3 in the training set
for an improved accuracy of the trained network.

By testing we established that \texttt{num\_train}=100000
for the original DSL that ships with PCCoder and \texttt{num\_train}=300000 for the DSL with {\em IFI} and {\em IFL}:

\begin{verbatim}
python -m scripts.gen_programs --num_train=300000 
--train_output_path=tt5nif_3 --test_output_path=tt5test 
--max_train_len=3 --num_workers=20 --cache=tt5nif_2
Loading program cache... 6628\6629
Generating programs of length 3 (current dataset size: 6629)
Generating programs... 300000\300000
Generating examples... 298578\300082 (remaining programs: 267583)
Discarding identical programs... 267393\267577
Finished generation. Total programs: 90214
Writing 90214 train programs to tt5nif_3
\end{verbatim}

The script thus generates (after more than 2 hours) a number of 90214 program examples of length at most 3
in the extended DSL. We proceed to further generate program examples of length at most 5, raising the cap
of raw generated programs to 600000:

\begin{verbatim}
python -m scripts.gen_programs --num_train=600000 
--train_output_path=tt5nif_5 --test_output_path=tt5test 
--max_train_len=5 --num_workers=20 --cache=tt5nif_3
Loading program cache... 90213\90214
Generating programs of length 4 (current dataset size: 90214)
Generating programs... 600100\600100
Generating examples... 597900\600446 (remaining programs: 483158)
Discarding identical programs... 482731\483104
Generating programs of length 5 (current dataset size: 158177)
Generating programs... 600100\600100
Generating examples... 596906\601198 (remaining programs: 388298)
Discarding identical programs... 387830\388043
Finished generation. Total programs: 174958
Removed 2525 programs
Writing 172433 train programs to tt5nif_5
\end{verbatim}

Thus (again after more than 2 hours) the script generates another 67963 program examples of length 4 and
another 14256 program examples of length 5, to a total of 172433 programs examples of length at most 5.

In comparison, for the original DSL it takes at least 4 hours to generate the set of 42730
program examples of length at most 3; we here had to opt for the smaller 100000 cap since
it was hardly attained due to the less expressive (original) DSL
\begin{verbatim}
python -m scripts.gen_programs --num_train=100000
 --train_output_path=t5nif_3 --test_output_path=t5test 
--max_train_len=3 --num_workers=20 --cache=t5nif_2
Loading program cache... 1585\1586
Generating programs of length 3 (current dataset size: 1586)
Generating programs... 100000\100000
Generating examples... 99308\100013 (remaining programs: 80603)
Discarding identical programs... 80411\80592
Finished generation. Total programs: 42730
Writing 42730 train programs to t5nif_3
\end{verbatim}
hence the size of the training set is slightly less than half the one for the extended DSL.
Going further for program examples of length at most 5 we get
\begin{verbatim}
python -m scripts.gen_programs --num_train=150000 
--train_output_path=t5nif_5 --test_output_path=t5test 
--max_train_len=5 --num_workers=20 --cache=t5nif_3
Loading program cache... 42729\42730
Generating programs of length 4 (current dataset size: 42730)
Generating programs... 150100\150100
Generating examples... 149123\150180 (remaining programs: 116402)
Discarding identical programs... 116276\116375
Generating programs of length 5 (current dataset size: 72480)
Generating programs... 150100\150100
Generating examples... 149206\150289 (remaining programs: 80598))
Discarding identical programs... 80453\80487
Finished generation. Total programs: 80494
Removed 423 programs
Writing 80071 train programs to t5nif_5
\end{verbatim}
hence a set of 80071 train programs, slightly less than half the number for the extended DSL
(adding 29750 of length 4 and 7591 of length 5). By raising the cap of \texttt{num\_train} to
200000 we get 95734 train programs (adding 42068 of length 4 and 10936 of length 5). It is
important not to raise the cap too much in order to allow the generation of programs of longer
length, as the generation script cannot produce more than a bit over 200000 train programs
altogether.

So if one is interested in programs of smaller length then it's ok to raise the
cap as much as possible, but otherwise it's better to proceed gradually.

The next step is to train a network from the generated training set, via
\begin{verbatim}
python -m scripts.train dataset model
\end{verbatim}
where \texttt{model} is the prefix name of the model file (a file \texttt{model\_i} is
saved after each training epoch $i$) and \texttt{dataset} is the previously generated
training dataset, e.g., \texttt{t5nif\_5} for the original DSL or \texttt{tt5nif\_5}
for the extended DSL. The number of epochs is given by a parameter \texttt{num\_epochs}
from the preamble of {\em scripts/train.py} which by default is 40.

For the original DSL the training of \texttt{m5} from \texttt{t5nif\_5} takes less than
4 minutes per epoch, hence roughly two hours to get \texttt{m5.39}. For the extended
DSL the training of \texttt{mm5} from \texttt{tt5nif\_5} takes 7 minutes per epoch,
hence four hours to get \texttt{mm5.39}.

At last we can proceed to test the program synthesis component, via

\begin{verbatim}
python -m scripts.solve_problems dataset result model 60 5 
--num_workers=8
\end{verbatim}
Here \texttt{dataset} is a file of inputs-output pairs, one for each line representing problems
to solve, given a timeout in seconds (here 60) and the maximal length of the sought program (here 5).
The script will display the number of problems that it could solve and the number of those who failed
to be solved (by the timeout). At the end it will give the number and the list of actual programs
in the \texttt{result} file.

We thus obtain, for the original DSL
\begin{verbatim}
python -m scripts.solve_problems t5test_6 res-test-6 
m5.39 3000 6 --num_workers=40
Solving problems... 100 (failed: 5)
Solved: 94\100: 94.0%
\end{verbatim}
meaning that programs were synthesized for 94 out of 100 sets of five inputs-output pairs for
test programs of length 6, in one case the timeout of 3000 seconds was overpassed and for
5 sets the script directly failed to produce a program.

When given test sets of programs of length at most 4 the success rate was 99\%.

The generalization capacity of this program synthesis method is illustrated when the network
trained on programs of length at most 5 yields a success rate of over 70\% on sets of test
programs of length 8 (in this case an astounding 80\%):
\begin{verbatim}
python -m scripts.solve_problems t5test_8 res-test-8 
m5.39 2000 8 --num_workers=40
Solving problems... 100 (failed: 19)
Solved: 80\100: 80.0%
\end{verbatim}

Even more, for a randomly generated test set of programs of length 14 the success rate
was of 50\%. In the case of the extended DSL we found by similar experiments that same
(minimal) percentages hold for the success rate, namely at least $99\%$ for programs
of length at most 5, at least $90\%$ for programs of length 6, at least $70\%$ for
programs of length 8 and $\approx 50\%$ for programs of length at most $14$.

\section{Conclusion}

The work of \cite{DeepCoder} brings a novel approach to program synthesis from inputs-output pairs, also called
programming-by-example (PBE) in that it uses a neural network to guide the search for programs compatible with
the input set of inputs-output examples. The authors manage to synthesize programs of length at most 5 in a DSL
that satisfy sets of 5 examples and their implementation is feasible for programs of length at most 8.

They use a trained neural network to predict an order on the program space and show how to use it to guide
search-based techniques that are common in the programming languages community
(enumerative search and SMT-based solver).
They thus bring a neural-guided solution to the Inductive Program Synthesis (IPS) problem.
While the approach of Balog et. al. works for short programs, the number of possible solutions grows exponentially
with program length, thus rendering the identification of the solution based on global properties infeasible.

The work of \cite{PCCoder} brings a step-wise approach to the program synthesis problem. Given the current state,
their neural network directly predicts the next statement, including both the function (operator) and parameters (operands).
They perform a beam search based on the network’s predictions, reapplying the neural network at each step. The more accurate
the neural network, the less programs one needs to include in the search before identifying the correct solution.
They also train a second network to predict the variables that are to be discarded, as the number of variables increases
with the program’s length and some of them may no longer be in use. Training the new network serves as an auxiliary task
that improves the generalization of the statement prediction network.

Both \cite{DeepCoder} and \cite{PCCoder} generate the train and test data in the same manner. First, generate random programs
from the DSL. Then, prune away programs that contain redundant variables, and programs for which an equivalent program
(possibly shorter) already exists in the dataset. Equivalence is approximated by identical behavior on a set of
inputs-output examples. Valid inputs for the programs are generated by bounding their output value to
the DSL’s predetermined range and then propagating these constraints backward through the program.

We successfully proved that the \cite{PCCoder} implementation can be extended to include
symbols for branching instructions that can be added to the DSL language with only a linear increase of 
the complexity of program search, but maintaining the accuracy of the search procedure.

We also extended the DSL language of DeepCoder/PCCoder with symbols for more complex functions for which we have
a (Python) implementation and saw whether the 1-instruction programs for such functions can be deduced by the
learning algorithm based on the two neural networks of PCCoder from limited sets of inputs-output correspondences.

We obtained a genuinely positive result for the case of a complex MNIST classification function.

\section{Future work}

We can identify three directions for future research concerning synthesis of programs from inputs-output examples.
First the DSL language may be extended with constructs allowing for cycling commands such as {\em repeat}; here
we prefer the variant of adding recursion to the language, by first alowing subroutines and then recursive
procedure calls. For this we can get inspiration from the recent paper \cite{PAGR} (see also \cite{EGNPS})
that achieved results in synthesizing complex sorting algorithms in the Neural Programming Architecture.

The second direction concerns experiments with different neural network topologies for the training part.
So far we only used the PCCoder \cite{GPCC} {\em Dense} network with a number of layers given as programmable
parameter. Originally 10 layers were used in \cite{GPCC} but we obtained the same results with 5 layers
and half the original \texttt{output\_size} and \texttt{growth\_size} parameters. Different topologies
may bring different results in terms of accuracy and generalizability.

Last but not least, a third direction is concerned with different program search algorithms. In \cite{GPCC}
beam search is used by default and Depth-First search is provided as alternative via the \texttt{search\_method}
parameter for the {\em solve\_problems} script. Recent work by Safari \cite{PPS} goes exactly in this
direction, with preliminary results indicating an increase in the speed of program search. 

\section{Acknowledgements}
This work was supported by the European Regional Development Fund and the Romanian Government through the
Competitiveness Operational Programme 2014--2020, project ID P\_37\_679, MySMIS code 103319, contract no. 157/16.12.2016.

\end{document}